\documentclass[aps,preprint,superscriptaddress]{revtex4-1}

\usepackage{amsfonts,amsmath,mathrsfs}
\usepackage{amssymb} 

\usepackage{bm} 

\usepackage{epsfig} 
\usepackage{float}

\usepackage{hyperref}
\usepackage{color}

\hypersetup{colorlinks=false}

\begin{document}

\title{Temporal variation of the spectrum of continuously-pumped random fiber laser. Phenomenological model.  }

\author{Y. Bliokh}
\affiliation{Department of Physics, Technion-Israel Institute of Technology, Haifa 32000, Israel}
\email{bliokh@physics.technion.ac.il}

\author{E. I. Chaikina}
\affiliation{\textbf{}Divisi\'{o}n de F\'{\i }sica Aplicada, 
	Centro de Investigaci\'{o}n Cient\' \i fica y de Educaci\'on Superior de Ensenada, 
	Km. 107 carretera Tijuana-Ensenada, Ensenada, B.\ C., 22860, M\'{e}xico}

\author{I. D. Vatnik}
\affiliation{Institute of Automation and Electrometry of the Siberian
	Branch of RAS, Ac. Koptyug ave 1 Novosibirsk 630090, Russia}
\affiliation{Novosibirsk State University,
	Pirogova str. 2, Novosibirsk 630090, Russia}

\author{D. V. Churkin}
\affiliation{Novosibirsk State University, Pirogova str. 2, Novosibirsk 630090, Russia}

\begin{abstract}
A temporal variation of a spectrum of excited modes in a continuously pumped
erbium-doped random fiber laser (RFL), based on randomly distributed Bragg gratings, is studied. Developed phenomenological theoretical model assumes hard excitation mechanism of the eigenmodes instability. The model explains qualitatively peculiarities of the spectrum variation, observed experimentally.
\end{abstract}

\maketitle

\section{Introduction}
Random lasers are complex photonic devices which rely on multiple scattering of light in non-homogeneous media with optical gain \cite{Wiersma}. A random laser was introduced in the late 1960s \cite{Letokhov} and was unambiguously demonstrated only in 1994 \cite{Lawandy}. Since then  random lasers have been thoroughly characterized in a diversity of systems, for example, biological materials \cite{Polson2004}, cold atoms \cite{Baudouin2013}, rare-earth-doped crystals \cite{Gomes2016} and optical fibers with different sources of random feedback \cite{Turitsyn2010,Guo2018}. 

The first demonstration of a random laser based on optical fiber employed a dye:TiO2 colloid in the hollow core of a photonic crystal fiber, constituting therefore a random fiber laser \cite{Matos}. This work was followed by the reports of  Liz\'arraga and coauthors \cite{Lizarraga} and Gagn\'e and Kashyap \cite{Gagne}, who exploited erbium-doped fibers with randomly distributed Bragg fiber gratings (FBG) as scattering elements.
 
Over last decades RFLs have been realized in  different configurations utilizing different amplification techniques.  RFLs demonstrate a number of promising properties, such as high efficiency generation, wide span tunability together with high robustness, narrow-band, cascaded laser, laser with polarized pumping and other \cite{Churkin}. The generation efficiency of a random DFB fiber laser can be superior to those of lasers with conventional cavity design due to the specific power distribution along the cavity \cite{Vatnik}.

The vastly different length scales on which random lasing may occur, and many different physical systems
in which they have been realized, have triggered the development of different theoretical approaches including Anderson localisation model, diffusion model, or Monte Carlo simulation \cite{Cao,Burin,Andreasen,Cao2,Soukoulis1,Soukoulis2,He2018}. Due to essentially multimode generation there are strong  fluctuations of the output radiation intensity\cite{Lima2017Intensity}, study of that may give insight to the processes underlying. Thus, instability of intensity may indicate on considerable nonlinear interactions of the light field with the active medium \cite{Cao3,Ohtsubo}, that can be considered in the paradigm of wave turbulence \cite{Churkin2015,Gonzalez2017}.
Fluctuations also can be found in the output spectra. This behavior has been shown to deviate from the Gaussian regime with weak fluctuations below the threshold switching to a L\'evy-like statistics above the onset of random laser emission \cite{Gomes2016}. This system was recently exploited as a photonic platform for studies of complex systems, such as spin - glass analogy through the observation of the replica-symmetry-breaking phase transition \cite{Lima2017LevyStatistics}. Moreover, it was shown recently \cite{Bittner} that utilizing cavities with complex spatial structure in a semiconductor medium may prevent the formation of self-organized structures such as filaments that are prone to modulations instabilities.

Although nonlinear interactions are usially assumed to result in a fast processes, it was shown that  multimode random fiber lasers with feedback due to a set of FBGs may possess strong deviations in the shape of the spectrum \cite{Lizarraga,ChaikinaSPIE2018} with large characteristic times approaching seconds. Such a behavior isn't described in the aforementioned models, while it's crucial for possible applications. 

While some features of the slow instabilities of the output radiation spectrum of a random fiber laser with feedback based on a set of fiber Bragg gratings  were presented in \cite{ChaikinaSPIE2018}, here we study in details  slow spectral dynamics and introduce a model that qualitatively explains the system behavior.

\section{Sample  and Experimental setup}
\begin{figure}[htb]
\centering\scalebox{0.7}{\includegraphics{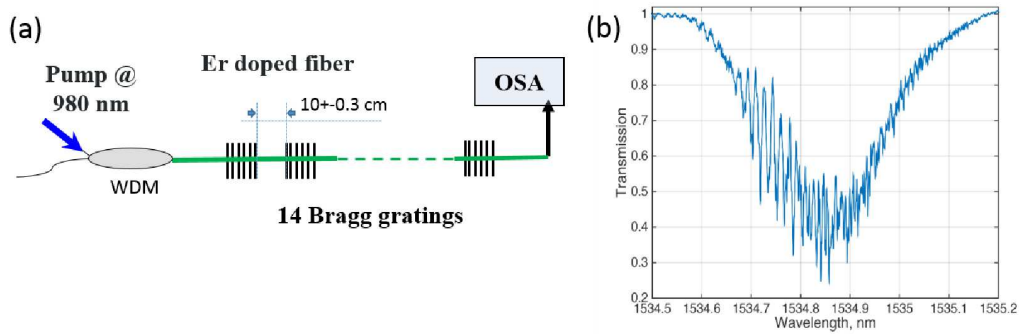}}
\caption{(a) Experimental setup of the laser with random feedback with FBG set. (b) Transmission spectrum of 14 randomly distributed Bragg gratings in passive regime. \label{Setup}}
\end{figure}

We built up a random fiber laser using a commercial Er/Ge co-doped single mode optical fiber Er304 made by INO (International Optics Institute). The core diameter is $D_{c} = 3.8 \, \mu$m,  and the maximum absorption is about $5.9\, $dB/m  at $1532 \,$ nm.
Fiber Bragg gratings (FBG) were inscribed in the fiber core by exposing with UV light from an intracavity frequency doubled Argon-ion laser ($\lambda _{r}=244 \, $nm) using a conventional mask technique \cite{Kashyap}. The $10 $ mm-long mask has a spatial period of $1059.8$ nm.
The length of each of the fabricated gratings $L_{g}$  was about $3 \,$mm, and the distances $L_{i}$ between two neighbouring gratings were randomly distributed in the range $10\pm 0.3 $ cm, Fig.~\ref{Setup}.  Total number of Bragg gratings is 14, and the total length of the structure is $L = N (L_{g}+L_{i})\approx  170 cm$. Transmission spectrum of the FBG set, has a number of peaks appeared at random places thus confirming Anderson localisation \cite{Bliokh}. The correlation length, as estimated considering a reflection coefficient of a single FBG of approximately 10\%, gives the value of $l_{loc}$ =10 gratings. The transmission spectrum of the fabricated laser in passive regime is presented at Fig.~\ref{Setup}(b). 

\section{Experimental results and analysis}

\begin{figure}[htb]
\centering\scalebox{0.3}{\includegraphics{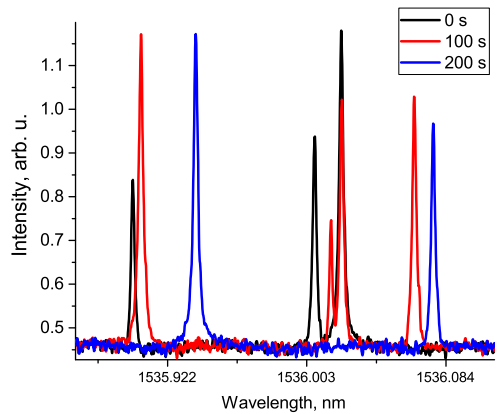}} \caption{The generated spectra of the laser measured at different times. Pump power is 680 mW.\label{spectra950mA}}  
\end{figure}

The  measurements of the spectrum of generation were performed using Finisar 1500s spectrum analyzer. It allows acquiring spectra with frequency of around 10 Hz and spectral resolution of 1.2 pm. We measured generation spectra for different levels $50-700 mW$ of the pump power. The spectrum consists of a number of narrow lines, each one corresponding to a localized mode built up in the cavity.
With that number of lines, a power in each line and position of the lines in the spectrum vary not only with pump power, but also with time \cite{Lizarraga, ChaikinaSPIE2018}, see Fig.~\ref{spectra950mA}.  The laser system under consideration may comprise as much as 124 modes \cite{Bliokh}, but  we observe only up to 10 generating modes. In order to study temporal instabilities in the spectrum we perform a number of long-term measurements at each pump power with acquisition rate of 10 Hz and duration of $500-900$ seconds.

\begin{figure}[htb]
\centering\scalebox{0.2}{\includegraphics{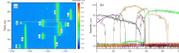}}  
\caption{a) Generation spectra of the laser measured at different times. Pump power is 700 mW.   b) Temporal dynamics of the intensities of the spectral lines depicted in (a). Note, that there are 9 lasing modes, however only from 2 to 4 (most often 3) modes are excited simultaneously. Letters in (a) and (b)	 mark the same spectral lines. \label{ExpSpectraDynamics} }
\end{figure}

The characteristic time of the generation of a particular mode may vary from few to hundreds seconds, see Fig.~\ref{ExpSpectraDynamics}a. Note that there is a strong temporal correlation between appearance of one spectral line and disappearance of the other one: modes are switching (see white lines in Fig.~\ref{ExpSpectraDynamics}a emphasizing switching). 
Once a mode disappears it can arise again later (see Fig.~\ref{ExpSpectraDynamics}a, line 'c'). Finally, generation build-up time is much smaller than the characteristic life-time for every mode (see Fig.~\ref{ExpSpectraDynamics}b).

\section{Theoretical model}

As it was mentioned above, the resonant frequencies of the disordered set of scatterers are randomly distributed in a certain interval, that allows one to suppose the absence of phase correlation between excited modes. Thus, one can use the simplest model which consists of the rate equations for inversion population density $n(t,x)$ and the modes intensities $I(x,t)$ \cite{Tang}:  

\begin{eqnarray}
\frac{da_\alpha(t)}{dt}=-a_\alpha(t)\left[\gamma_{diss}^{(\alpha)}-G\intop_0^L dx\, n(t,x)F_\alpha(x) \right]\label{eq1}\\ 
\frac{dn(t,x)}{dt}=\frac{n_0-n(t,x)}{t_{rel}}-Gn(t,x)\sum_\alpha a_\alpha(t)F_\alpha(x).  \label{eq2}
\end{eqnarray} 
Here the intensity $I_\alpha (x,t)$ of the $\alpha$-th mode is presented in the form $I_\alpha(x,t)=a_\alpha(t)F_\alpha(x)$, spatial distribution of the intensity along the fiber of the length $L$ is described by the function $F_\alpha(x)$. $\gamma_{diss}^{(\alpha)}$ describes the QNMs intrinsic energy loss (dissipation, leakage), $t_{rel}$  is the relaxation time of the inverse population, $G$ is the gain coefficient, $n_0$ is the equilibrium population in the absence of modes field.

Let us rewrite Eqs.~(\ref{eq1}) and (\ref{eq2}) in dimensionless form:
\begin{eqnarray}
\frac{db_\alpha}{d\tau}=-b_\alpha\left[\gamma^{(\alpha)}-g\intop_0^1d\xi\,\eta\tilde{F}_\alpha   \right]\label{eq3}\\
\frac{d\eta}{d\tau} =(1-\eta)-g\eta \sum_\alpha b_\alpha\tilde{F_\alpha}. \label{eq4}
\end{eqnarray}
Here $\tau=t/t_{rel}$, $\xi=x/L$, $\eta=n/n_0$, $b_\alpha=a_\alpha/L$, $g=Gn_0t_{rel}L$,  $\gamma^{(\alpha)}=\gamma_{diss}^{(\alpha)}t_{rel}$. The eigenfunctions $\tilde{F_\alpha}(\xi)$ are normalized: 
\begin{equation}
\label{eq4a}
\intop_0^1d\xi\tilde{F_\alpha}(\xi)=1
\end{equation}

First let us estimate the threshold value of the gain coefficient $g$, which correspond to the mode excitation. In a random medium QNMs are localized states, whose spatial structure is defined by the localization length, $\xi_\ell$, and position of the localization center, $\xi_\alpha$: $\tilde{F}(\xi)\propto \exp(-|\xi-\xi_\alpha|/\xi_\ell)$. Assuming that $\xi_\ell\ll 1$ and considering only QNMs which are placed rather far from the fiber ends, (at the distance of the order of $\xi_\ell$ or more), one can extend limits in the normalization integral Eq.~(\ref{eq4a}), so that $\tilde{F}_\alpha(\xi)\simeq(1/2\xi_\ell)\exp(-|\xi-\xi_\alpha|/\xi_\ell)$. The QNMs which are placed closer to the fiber ends can be excluded from the consideration because their dissipation coefficient  $\gamma_{diss}^{(\alpha)}$ is large and these modes have either large excitation threshold or even are not excited at all.

Let assume, that only one mode is excited. The amplitude of this mode is defined by steady-state solutions of Eqs.~(\ref{eq3}) and (\ref{eq4}). It follows from Eq.~(\ref{eq4}):
\begin{equation}\label{eq5}
\eta=\frac{1}{1+gb\tilde{F}}.
\end{equation}
Substitution Eq.~(\ref{eq5}) int Eq.~(\ref{eq3}) leads to equation, which defines the steady-state amplitude of unstable mode:
\begin{equation}\label{eq6}
\gamma=g\intop_0^1d\xi\,\frac{\tilde{F}}{1+gb\tilde{F}}.
\end{equation}
Integral in Eq.~(\ref{eq6}) can be easily evaluated ($\xi_\ell\ll 1$ !),
\begin{equation}\label{eq7}
\intop_0^1d\xi\,\frac{\tilde{F}}{1+gb\tilde{F}}=\frac{2\xi_\ell}{gb}\ln(1+gb/2\xi_\ell),
\end{equation} 
and Eq.~(\ref{eq6}) can be presented as follows::
\begin{equation}\label{eq8}
\gamma=\frac{2\xi_\ell}{b}\ln(1+gb/2\xi_\ell).
\end{equation}
Equation (\ref{eq8}) has solution only when
\begin{equation}\label{eq9}
g\geq g_{th}=\gamma.
\end{equation}

For a localized state in a dissipationless medium the dissipation coefficient $\gamma_{diss}$ is defined by a wave leakage from the sample edges: $\gamma_{diss}=\gamma_{leak}=-(v_{g}/L)\ln(R_1R_2)$, where $v_{g}$ is the wave group velocity into the medium, and $R_{1,2}$ are the reflection coefficients from the walls of the effective cavity formed due to the Anderson localization. Thus,
\begin{eqnarray}
\label{eq10}
\gamma_{leak}\propto\left|\ln\left(R_1R_2\right)\right|=\left|\ln\left[(1-T_1)(1-T_2) \right]\right|\simeq T_1+T_2\simeq e^{-x_0/\ell}+e^{-(L-x_0)/\ell}\nonumber\\
=2 e^{-L/2\ell}\cosh\left[\left(L/2-x_0\right)/\ell\right].
\end{eqnarray}
Here $T_{1,2}=1-R_{1,2}\ll 1$ are transmission coefficient of the effective cavity walls.
In the dimensionless form
\begin{equation}
\label{eq11}
\gamma^{\alpha}=\gamma^{(\alpha)}_{leak}=\gamma^{(\alpha)}_{leak 0}\cdot e^{-1/2\xi_\ell}\cosh\left[\left(1/2-\xi_\alpha\right)/\xi_\ell\right].
\end{equation}

\begin{figure}[htb]
\centering \scalebox{0.3}{\includegraphics{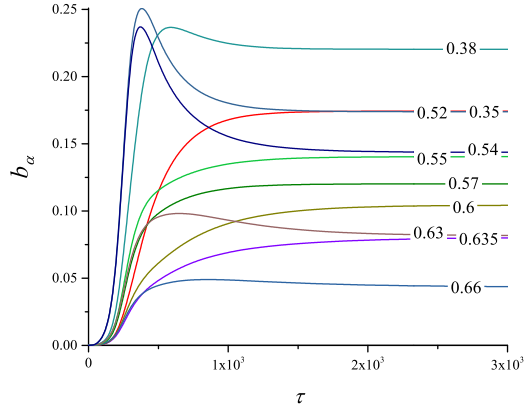}} \caption{Excited modes of random laser. Total number of modes $N_{mod}=30$, only 10 modes are excited. Centers of the localized modes, $\xi_{\alpha}$, are randomly distributed along the fiber. Curves marked by the $\xi_{\alpha}$ values.    \label{Fig3}} 
\end{figure}

Example of numerical solution of Eqs.~(\ref{eq3}) and (\ref{eq4}) with $\gamma^{(\alpha)}$ from Eq.~(\ref{eq11}) is shown in Fig.~\ref{Fig3}. Note that the number of lazing modes $N_{las}=10$ is smaller than the total number of modes $N_{mod}=30$. 

Dimensionless localization length in this example is equal to 0.1, $\xi_\ell=0.1$. All the lazing modes are located at the distances smaller than $\xi_\ell$ from the fiber middle, $|\xi_{\alpha}-0.5|<\xi_\ell$, and there are no another modes in this region. Despite the fact that all these modes are more or less overlapped, the competition for inverse population does not lead to the total  suppression of one mode by the other one. The only reason why namely these modes are excited is their relatively small dissipation coefficients, defined by Eq.~(\ref{eq11}) and satisfying condition Eq.~(\ref{eq9}).

Because the competition itself cannot explain observed experimentally switching between lazing modes, the model should be modified. It was mentioned above that when some parameters of excited mode vary with time in such a way that the excitation condition Eq.~(\ref{eq9}) is violated, then the suppressed mode can be excited. This variation of the mode parameters should be different for the suppressed and dominant modes, otherwise condition Eq.~(\ref{eq9}) will be violated for both modes, i.e., the initially suppressed mode will never be excited. Thus, the rate of change of the mode's parameters depends, either directly or indirectly,  on the mode amplitude.

Determination of physical reasons that lead to the parameter variation is out of scope of the paper. The aim of this research is to show that this behavior can, in principle, explain experimental data. Therefore we will consider further the simplest case when only one parameter in Eq.~(\ref{eq9}), namely $\gamma^{(\alpha)}$, is varied.

In general, the dissipation coefficient $\gamma^{(\alpha)}$ includes leakage and volumetric dissipation:
\begin{equation}\label{eq11a}
\gamma^{(\alpha)}=\gamma^{(\alpha)}_{leak}+\gamma^{(\alpha)}_{vol}.
\end{equation}
Let us assume that the dissipation coefficient $\gamma^{(\alpha)}_{vol}$ depends on a ``temperature'' $T_\alpha$ which, in its turn, is defined by ``heating'' of an effective resonator, associated with the given mode $\alpha$, by the excited field with amplitude $b_\alpha$:
\begin{eqnarray}
\gamma^{(\alpha)}_{vol}=\gamma^{(\alpha)}_{vol0}T_\alpha, \label{eq12}\\
\frac{dT_\alpha}{d\tau}=\kappa_+ b_\alpha-\kappa_-\left(T_\alpha-1\right).\label{eq13}
\end{eqnarray}
Here $\kappa_\pm$ are the coefficients. The first term in the right-hand-side of Eq.~(\ref{eq13}) describes  ``heating'', the second one describes relaxation to the environment's  ``temperature'' $T_0=1$.

The volumetric dissipation can, in principle, be frequency-dependent, similarly to a frequency-dependent gain. This justify introduction of individual ``temperatures'' $T_\alpha$ for different resonators.

\begin{figure}[htb]
\centering \scalebox{0.3}{\includegraphics{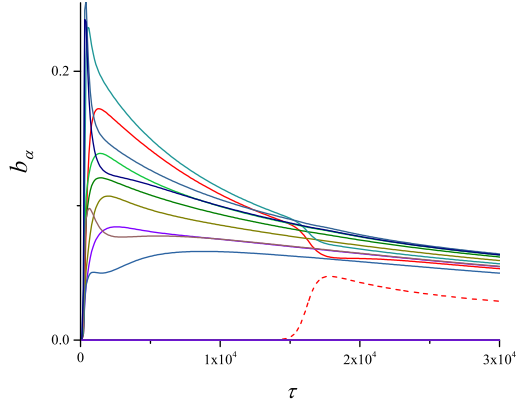}} \caption{Temporal evolution of excited modes with ``temperature''-dependent dissipation coefficients. Amplitudes of the excited modes decrease due to the ``heating''. The ``consumption'' of the inverse population decreases that allows initially suppressed mode (dashed curve) to be excited.       \label{Fig4}} 
\end{figure}

Example of numerical solution of Eqs.~(\ref{eq3}) and (\ref{eq4}) with $\gamma^{(\alpha)}$ defined by  Eqs.~(\ref{eq11a}), (\ref{eq12}), (\ref{eq13})  and the same values of other parameters is shown in Fig.~\ref{Fig4}.

Interaction between the modes through the competition for inverse population is visible in Fig.~\ref{Fig4} at time $\tau\simeq 1.5\cdot 10^4$, but the switching between modes is absent in this model also. 

In order to understand why the previous models do not lead to the desired result, let us present Eq.~(\ref{eq3}) for one mode in the form
\begin{equation}\label{eq14}
\frac{db}{d\tau}=-b\left[\Gamma_{diss}(b)-\Gamma_{gain}(b)\right]\equiv b\Phi(b).
\end{equation}
The steady-state solutions are $b=0$ and positive zeros of $\Phi(b)$, if they exist.

\begin{figure}[htb]
\centering \scalebox{0.3}{\includegraphics{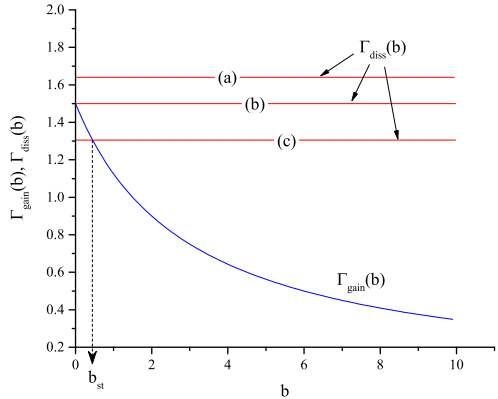}} \caption{Soft excitation. (a) -- no excitation; (b) -- threshold; (c) -- excitation, $b_{st}$ is the steady-state amplitude.  \label{Fig5}} 
\end{figure}

In the models above function $\Gamma_{diss}$ does not depend \textit{explicitly} on the amplitude $b$;
\begin{equation}\label{eq14a}
\Gamma_{diss}(b)=\gamma_{leak}+\gamma_{vol}(T).
\end{equation}
In contrast, $\Gamma_{gain}(b)$ depends on the amplitude: 
\begin{equation}\label{eq14b}
\Gamma_{gain}(b)=\frac{2\xi_\ell}{b}\ln(1+gb/2\xi_\ell)
\end{equation}
[see Eq.~(\ref{eq8})]. $\Gamma_{gain}(b)$ is monotonically  decreasing function whose maximal value in the region $b\geq 0$ is $\Gamma_{gain}(0)=g$, so that the equation $\left.\Phi(b)\right|_{b\geq 0}=0$ has solution when the condition Eq.~(\ref{eq9}) is satisfied. The smaller is the excess over threshold, the smaller is the steady-state mode's amplitude (see Fig.~\ref{Fig5}).  The same is correct when many modes can be excited simultaneously. During the transient process the threshold conditions for any mode are established self-consistently through competition for inverse population. What is important that amplitudes of the modes are changed smoothly, without sharp ``jumps'', because the amplitude tends to zero when the mode parameters approach threshold values. 
The same behavior is inherent in the model with ``temperature''-dependent dissipation coefficients [see eqs.~(\ref{eq12}), (\ref{eq13})]. 

\begin{figure}[htb]
\centering \scalebox{0.3}{\includegraphics{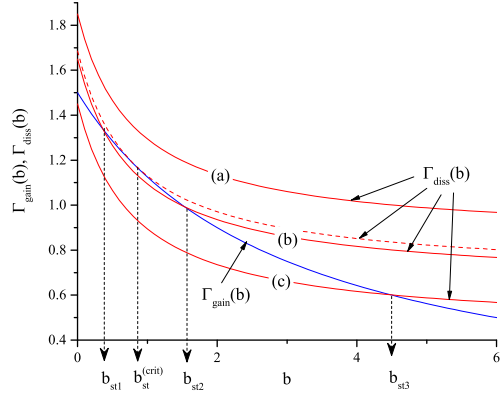}} \caption{Hard excitation. (a) -- no excitation. (b) -- the mode is excited only when its initial amplitude exceeds $b_{st1}$. The steady-state amplitude is $b_{st}=0$ or $b_{st2}$. (c) -- Excitation from small initial amplitude, $b_{st3}$ is the steady-state amplitude.    \label{Fig6}} 
\end{figure}

The situation can be changed drastically when the function $\Gamma_{diss}$ also depends \textit{explicitly} on amplitude $b$. In this case equation $\Gamma_{disss}(b)=\Gamma_{gain}(b)$ with monotonically-decreasing function $\Gamma_{diss}(b)$ can get more than one solution, as it is shown in Fig.~\ref{Fig6}. 

\begin{figure}[htb]
\centering \scalebox{0.3}{\includegraphics{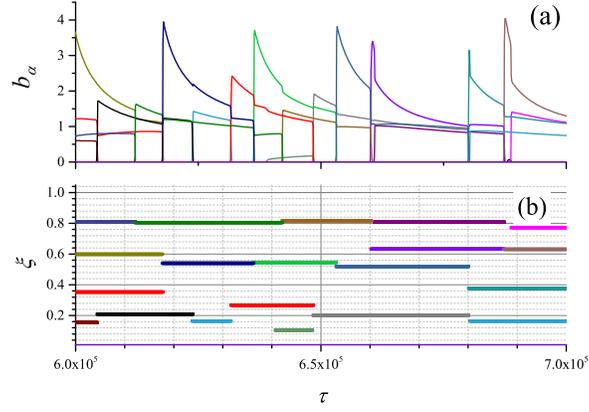}} \caption{ Random laser model with hard excitation. (a) -- Amplitudes $b_\alpha(\tau)$ of lazing modes. Total number of modes $N_{mod}=30$. Note that only 3 -- 4 modes are excited simultaneously. (b) -- Position $\xi_\alpha$ of excited lasing mode.  \label{Fig7}} 
\end{figure}

Figures ~\ref{Fig5} and ~\ref{Fig6} demonstrate the difference between soft (Fig.~\ref{Fig5}) and hard (Fig.~\ref{Fig6}) self-excitation mechanisms: in the first case the  oscillation amplitude varies smoothly under the system parameters slow variation, while the second mechanism is characterized by step-wise appearance/disappearance of oscillations and hysteresis -- oscillations are excited and suppressed at different values of the parameters. The step-like dependence of the steady-state amplitude on the parameters can explain the modes switching phenomenon. Indeed, presented in Fig.~\ref{Fig7} results of numerical solution of Eqs.~(\ref{eq3}), (\ref{eq4}) with ``temperature''-dependent non-linear dissipation coefficients $\gamma^{(\alpha)}(b;T))$, similar to $\Gamma_{diss}(b)$ depicted in Fig.~\ref{Fig6}, demonstrates switching between modes.  

A comparison between figures \ref{ExpSpectraDynamics}a,b and figures \ref{Fig7}a,b shows that the developed model of random laser with hard excitation possesses all essential property of observed experimentally radiation. 

It is worth noting that at the beginning of lazing, when the pump turn on, almost all potentially unstable modes are excited simultaneously. Later, after transit process, only several oscillating modes, whose number grows with increasing of the pump value,  can be observed at one time. It means that the modes switching is unique to lasers with long-time pumping. 

\section{Conclusion}

Temporal evolution of the spectrum  of continuously-pumped random fiber laser, based on the Er-doped fiber with randomly distributed Bragg gratings, has been studied experimentally and in the frame of developed phenomenological mathematical model. The simplest model which consists of the rate equations for inversion population density and the modes intensities \cite{Tang} is used as a basis and modified in order to fit experimentally observed peculiarities of the spectrum temporal evolution. It is shown that incorporation of a saturated absorption into the model and assumption, that there is a certain slow process (like, e.g., heating) caused degradation of Q-factor of the excited mode, can change scenario of the oscillations excitation.  Instead of the soft excitation, which is characterized by continuous growth of the mode amplitude with the pump excess over the threshold value, the hard excitation mechanism is responsible for relatively fast step-like appearance and disappearance of the radiated spectral line.

Numerical studies of the proposed model demonstrate features peculiar to the spectrum dynamics, listed below:       
\begin{enumerate}
	\item[(i)] {the number of simultaneously observed lines, $N_{las}$, is much smaller than the expected number of eigenmodes, $N_{mod}$,  in this frequency interval;}
	\item[(ii)] {any spectral line exists during only some time interval (usually several tens of seconds);}
	\item[(iii)] {there is a strong temporal correlation between appearance of one spectral line and disappearance of the other one (modes switching);
	\item[(iv)] one time disappeared emission line could arise again with the same frequency;}
	\item[(v)] {the characteristic time scale of the spectral line switching is much smaller  than the characteristic ``life time''}. 
\end{enumerate}
 
In spite that the developed mathematical model is able to reconstruct qualitatively the experimental data, the further investigations, both experimental and theoretical, are essential for revealing of concrete physical processes, which can justify the model's basis assumptions.

\section*{Funding}
The work was partially supported by  Newton International Fund of The Royal Society through the grand NI150255. I. D. V. acknowledges support from the Russian Foundation for Basic Research (16-32-60184,17-02-00929/17).

\end{document}